\documentclass[twocolumn,showpacs,preprintnumbers,amsmath,amssymb]{revtex4}
\usepackage{graphicx}% Include figure files
\usepackage{dcolumn}% Align table columns on decimal point
\usepackage{bm}% bold math

\begin{document}

\preprint{APS/123-QED}

\title{Mapping Monte Carlo to Langevin dynamics: A Fokker-Planck approach}% Force line breaks with \\

\author{X. Z. Cheng}
\author{M. B. A. Jalil}
\affiliation{%
Department of Electrical and Computer Engineering, National University of Singapore, 4 Engineering Drive 3, 117576, Singapore
}%

\author{Hwee Kuan Lee}
\altaffiliation[Current Address: ]{Data Storage Institute, 5
Engineering Drive 1, DSI  Building, Singapore, 117608}
%\affiliation{
%Department of Physics, Tokyo Metropolitan University, 1-1 Minami-Osawa, Hachioji-shi, Tokyo, Japan 192-0397\\
%Data Storage Institute, 5 Engineering Drive 1, DSI  Building, Singapore, 117608
%}%

\author{Yutaka Okabe}
\affiliation{
Department of Physics, Tokyo Metropolitan University,
1-1 Minami-Osawa, Hachioji-shi, Tokyo, Japan 192-0397
}%

\date{\today}% It is always \today, today,
             %  but any date may be explicitly specified

\begin{abstract}
We propose a general method of using the Fokker-Planck equation
(FPE) to link the Monte-Carlo (MC) and the Langevin micromagnetic
schemes. We derive the drift and diffusion FPE terms corresponding
to the MC method and show that it is analytically equivalent to the
stochastic Landau-Lifshitz-Gilbert (LLG) equation of Langevin-based
micromagnetics. Subsequent results such as the time quantification
factor for the Metropolis MC method can be rigorously derived from
this mapping equivalence.  The validity of the mapping is shown by
the close numerical convergence between the MC method and the LLG
equation for the case of a single magnetic particle as well as
interacting arrays of particles. We also found that our Metropolis
MC is accurate for a large range of damping factors $\alpha$, unlike
previous time-quantified MC methods which break down at low
$\alpha$, where precessional motion dominates.
\end{abstract}

% PACS, the Physics and Astronomy
                             % Classification Scheme.
%\keywords{Suggested keywords}%Use showkeys class option if keyword
                              %display desired
\maketitle

With the rapid advance of computing resources, Monte Carlo (MC)
methods have become a powerful tool in many fields ranging from the
physical sciences to finance and sociology
\cite{landau,finance,stauffer}. The flexibility of Monte Carlo is
due to its abstract formalism which can be realized in an almost
infinite number of ways. Increasingly, MC methods are being
implemented in the stochastic micromagnetic modeling of magnetic
nanostructures \cite{fidler}.  Stochastic micromagnetic modeling has
important practical implications, e.g. in predicting the storage
lifetime of hard-disk magnetic media \cite{neel,brown59}.
Traditionally, the dynamics of magnetic moments are also modeled in
the Langevin scheme, using the stochastic Landau-Lifshitz-Gilbert
(LLG) equation \cite{brown63}. Langevin-based micromagnetics
constitutes a formidable computational method, because of its ease
of use and close correspondence to actual experimental data in
previous literatures \cite{coffeyprl}. However, it has certain
limitations which may be overcome by MC methods. For instance, MC
methods can accommodate the long-time magnetization relaxation
dynamics of large-scale arrays of magnetic grains
\cite{novotny,kol,hk05}, which is practically unfeasible to be
modeled using the stochastic LLG equation. On the other hand, MC
schemes have the drawback of having its time calibrated in MC steps,
instead of physical time units. Thus, both MC and the Langevin
approach are useful computational methods in micromagnetics, with
complementary strengths and drawbacks. Hence, it is very important
to devise a general way of mapping one method to the other, and vice
versa.

Early efforts to link MC to LLG were done by Nowak \emph{et al.}
\cite{nowak}. They focused on deriving a time quantification factor
to relate one MC step (MCS) to real physical time unit used in the
LLG equation. Recently, we also proposed another time-quantifiable
MC (TQMC) method which involves the determination of macroscopic
density of states, and the use of the Master equation for time
evolution. This method is applicable in simulating extremely long
time magnetization reversal process \cite{ourprb}. The effect of
precession on Nowak's time-quantification was investigated by
Chubykalo \emph{et al.} \cite{chubykalo}. They concluded that
Nowak's time quantification of MC breaks down in the low damping
case, in presence of an oblique external field, due to the influence
of (athermal) precessional motion.

In this paper, our approach is more thorough and fundamentally
different from all previous works. We propose a systematic way of
using the Fokker-Planck equation (FPE) to map MC to LLG dynamical
equation. The physical background of using FPE to describe
stochastic dynamics has been well established \cite{reif}, e.g. in
the case of a particle under the influence of a one-dimensional
potential \cite{kikuchi}. The outline of our scheme is as follows.
We consider a single isolated particle and then generalize to an
interacting array of particles. First, we develop a MC method that
has the stochastic dynamics of the LLG equation. This method is a
hybrid Metropolis MC scheme which combines the random displacement
of spins about a cone and a suitably-sized precessional step. The
random displacement models the thermal fluctuation and the
precessional step accounts for the precessional term in the LLG
equation. From the drift-diffusion picture of the general
Fokker-Planck equation, we can then obtain the FP coefficients of
this hybrid Metropolis MC method, and compare them with previously
obtained FP coefficients of the stochastic LLG equation. The
comparison shows an exact equivalence of the two FPEs in the limit
of small MC step size. From this comparison, we obtain the time
quantification factor to relate one MC step to real time unit and
the required step size for the precession, which allows us to map
the MC results to that of the LLG equation. Generalization to an
array of interacting particles will be shown numerically in the
latter part of this paper.

We consider an isolated single domain magnetic particle whose moment
can be represented by a Heisenberg spin with an easy axis anisotropy
\cite{brown63}. To describe the dynamics of a Heisenberg spin, it is
convenient to use the spherical coordinate system. The FPE in
spherical coordinates $\theta$ and $\varphi$ can be written in form
as:
%
%   Eq: general form of FPE
%
\begin{eqnarray}
\frac{\partial P}{\partial t}
&=&
-\frac{\partial}{\partial\theta}\left(A_{\theta}\cdot P\right)-\frac{\partial}{\partial\varphi}\left(A_{\varphi}\cdot P\right)+\frac{1}{2}\frac{\partial^2}{\partial\theta\partial\varphi}\left(B_{\theta\varphi}\cdot P\right)\nonumber\\
& &
+\frac{1}{2}\frac{\partial^2}{\partial\theta^2}\left(B_{\theta\theta}\cdot
P\right)+\frac{1}{2}\frac{\partial^2}{\partial\varphi^2}\left(B_{\varphi\varphi}\cdot
P\right)
\label{eq:one}
\end{eqnarray}
$P=P(\theta,\varphi,t)$ is the probability density of the moment
orientation. $A$ and $B$ are the so-called drift and diffusion
coefficients respectively, defined as the ensemble mean of an
infinitesimal change of $\theta$ and $\varphi$ with respect to time
\cite{risken}.

The reduced stochastic LLG dynamical equation can be written as:
%
%   Eq: reduced form of stochastic LLG equaion
%
\begin{equation}
\frac{d{\bf m}}{dt}=-\frac{\gamma_0 H_k}{1+\alpha^2}{\bf
m}\times\left[\left({\bf h}+{\bf h}_t\right)+\alpha\cdot{\bf
m}\times\left({\bf h}+{\bf h}_t\right)\right]
\label{eq:two}
\end{equation}
where ${\bf m}$ is the magnetic moment unit vector, $\alpha$ and
$\gamma_0$ are the damping and gyromagnetic constant respectively,
${\bf h}$ is the effective field normalized by the anisotropy field
$H_k=2K_u/\mu_{0} M_s$, where $K_u$ is the anisotropy constant,
$\mu_0$ is the magnetic permeability and $M_s$ is the saturation
magnetization. The thermal field ${\bf h}_t$ is introduced by Brown
\cite{brown63} as a white noise term. The FPE corresponding to the
LLG equation has been derived by Brown \cite{brown63}, and its
factors are as follows:
%
%   Eq: FPE factors for LLG from Brown
%
\begin{eqnarray}
A^{LLG}_{\theta}&=&-h'\frac{\partial
E}{\partial\theta}-g'\frac{1}{\sin\theta}\frac{\partial
E}{\partial\varphi}+k'\cot\theta\nonumber
\\
A^{LLG}_{\varphi}&=&g'\frac{1}{\sin\theta}\frac{\partial
E}{\partial\theta}-h'\frac{1}{\sin^2\theta}\frac{\partial
E}{\partial\varphi}\nonumber
\\
B^{LLG}_{\theta\theta}&=&2k'\label{eq:three}
\\
B^{LLG}_{\varphi\varphi}&=&\frac{1}{\sin^2\theta}2k'\nonumber
\\
B^{LLG}_{\theta\varphi}&=&0\nonumber
\end{eqnarray}
where in above equations,
$h'=\frac{\alpha\gamma_0}{\mu_{0}VM_{s}(1+\alpha^{2})}$,
$g'=h'/\alpha$, $k'=h'/\beta$, $E$ is the total energy
\cite{brown63,nowak}, $V$ is the volume of the particle and
$\beta=(k_{B}T)^{-1}$, $k_B$ is the Boltzmann constant and $T$ is
the temperature in Kelvin.

We will now derive the FPE corresponding to our Monte Carlo method.
For the MC method, we choose with probability $q$, to displace the
magnetic moment within a small cone centered at the original
magnetization direction, and with probability $(1-q)$ to perform a
rejection-free precession about an effective field. For the
displacement about a cone, we pick a random vector lying within a
sphere of radius $R$ to the original magnetic moment and then
normalize the resulting vector. The precessional step vector, i.e.
the displacement of magnetic moment due to precession, is
${\Delta}{\bf m}=-\Phi\cdot{\bf m}\times{\bf h}$, where $\Phi\ll 1$
is a precesional step size to be determined. The probability $q$ is
chosen to be $1/2$, which yields a near-optimal balance of
efficiency and accuracy of our simulation.
\begin{figure}
\includegraphics[width=0.40\textwidth]{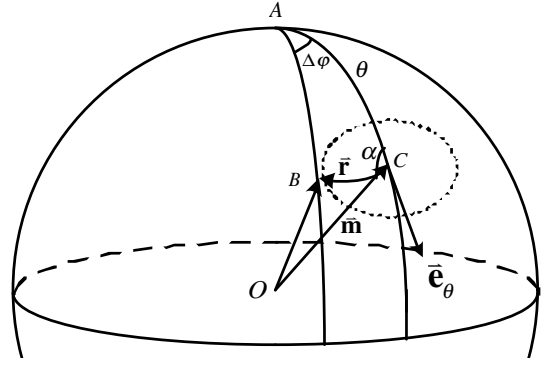}% Here is how to import EPS art
\caption{\label{fig:sphere} Diagram of random walk step of length
$r$ and angle $\alpha$ to $\vec{\bf e}_{\theta}$ which define a
spherical triangle ABC.}
\end{figure}

To calculate the FP coefficient $A^{MC}_{\theta}$ for the MC method,
we obtain the ensemble mean of a small change of $\theta$ in
\emph{one Monte Carlo step}. Contributions from random walk and
precessional step are
$A^{MC}_{\theta}=\left<\Delta\theta\right>_{\mbox{\small rand}}
/2+\Delta\theta_{\mbox{\small prec}}/2$, where the angle brackets
denote the ensemble average.

We first calculate the $\left<\Delta\theta\right>_{\mbox{\small
rand}}$, where the angular displacement is defined by two random
variables $r$ and angle $\alpha$, as shown in Fig.~\ref{fig:sphere}.
After some geometrical analysis, we obtain:
%
%   Eq: Sphere triangle solution
%
\begin{eqnarray}
\Delta\theta_{\mbox{\small rand}}&=&-\cos\alpha\cdot
r+\frac{1}{2}\cot\theta\sin^2\alpha\cdot r^2+O\left(r^3\right)
\label{eq:four}
\\
\Delta\varphi_{\mbox{\small rand}}
&=&\frac{1}{\sin\theta}\sin\alpha\cdot
r+\frac{1}{2}\frac{\cot\theta}{\sin\theta}\sin 2\alpha\cdot
r^2+O\left(r^3\right) \label{eq:five}
\end{eqnarray}
Next, we require the probability for the displacement vector to be
of size $r\ (r<R)$ and angle $\alpha$ with respect to $\vec{{\bf
e}}_{\theta}$. This probability is given by Nowak \emph{et al.}
\cite{nowak} as $p(r)=3\sqrt{R^2-r^2}/2\pi R^3$. Based on the
heat-bath Metropolis MC scheme, the acceptance rate is
%
%   Eq: Metropolis MC accepting rate
%
\begin{eqnarray}
A\left(\Delta E\right)&=&
%\frac{1}{1+\exp\left(\beta\Delta E\right)}\nonumber\\
1/\left(1+\exp\left(\beta\Delta E\right)\right)\nonumber\\
&\approx&\frac{1}{2}\left(1-\frac{1}{2}\beta\left(\frac{\partial
E}{\partial\theta}\Delta\theta+\frac{\partial
E}{\partial\varphi}\Delta\varphi\right)\right) \label{eq:six}
\end{eqnarray}
where $\Delta E$ is the energy change in the random walk step. Thus,
integrating over the projected surface of Fig.~\ref{fig:sphere}, one
obtains $\left<\Delta\theta\right>_{\mbox{\small rand}}$:
%
%   Eq: The total contribution to the FPE from random walk
%
\begin{eqnarray}
\left<\Delta\theta\right>_{\mbox{\small rand}}&=&\int_{0}^{2\pi}d\alpha\int_{0}^{R}(rdr)\Delta\theta\cdot p(r)\cdot A(\Delta E)\nonumber\\
&=&\frac{R^2}{20}\left(\cot\theta-\beta\frac{\partial
E}{\partial\theta}\right)+O(R^3) \label{eq:seven}
\end{eqnarray}

Next we calculate the other contribution from the precessional step
$\Delta\theta_{\mbox{\small prec}}$:
%
%   Eq: the contribution of precessional step to the FPE
%
\begin{eqnarray}
\Delta\theta_{\mbox{\small prec}}&\cong&{\bf e}_{\theta}\cdot\left(-\Phi\cdot{\bf m}\times{\bf h}\right)=\Phi\cdot({\bf e}_\varphi\cdot{\bf h})\nonumber\\
&=&-\frac{1}{\sin\theta}\frac{\Phi}{2K_{u}V}\frac{\partial
E}{\partial \varphi} \label{eq:eight}
\end{eqnarray}

In the above derivation, we have used the vector identity ${\bf
a}\cdot({\bf b}\times{\bf c})=({\bf a}\times {\bf b})\cdot{\bf c}$
and ${\bf h}=-(\bigtriangledown_{\bf m} E)/2K_{u}V$. Using
Eqs.~(\ref{eq:seven}) and (\ref{eq:eight}), $A^{MC}_{\theta}$
becomes,
%
%   Eq: FPE factor for MC method
%
\begin{equation}
A_{\theta}^{MC}=\frac{R^2}{40}\left(\cot\theta-\beta\frac{\partial
E}{\partial\theta}\right)-\frac{1}{\sin\theta}\frac{\Phi}{4K_{u}V}\frac{\partial
E}{\partial\varphi}+O(R^3)
\label{eq:nine}
\end{equation}

The other FPE factors can be obtained with the same procedure.
%
%   Eq: Other FPE factors for MC method
%
\begin{eqnarray}
A_{\varphi}^{MC}&=&-\frac{1}{\sin^2\theta}\frac{R^2}{40}\beta\frac{\partial
E}{\partial\varphi}+\frac{1}{\sin\theta}\frac{\Phi}{4K_{u}V}\frac{\partial
E}{\partial\theta}+O(R^3)\nonumber
\\
B_{\theta\theta}^{MC}&=&\frac{R^2}{20}+\frac{1}{2}\left(\frac{1}{\sin\theta}\frac{\Phi}{2K_{u}V}\frac{\partial
E}{\partial\varphi}\right)^2+O(R^4)\nonumber
\\
B_{\varphi\varphi}^{MC}&=&\frac{1}{\sin^2\theta}\frac{R^2}{20}+\frac{1}{2}\left(\frac{1}{\sin\theta}\frac{\Phi}{2K_{u}V}\frac{\partial
E}{\partial\theta}\right)^2+O(R^4)\nonumber
\\
B_{\theta\varphi}^{MC}&=&-\left(\frac{1}{\sin\theta}\frac{\Phi}{2K_{u}V}\right)^2\frac{\partial
E}{\partial\theta}\frac{\partial E}{\partial\varphi}+O(R^3)
\label{eq:ten}
\end{eqnarray}

We can now compare the FPE factors corresponding to the Langevin
(LLG) equation in Eq.~(\ref{eq:three}), with those of the Metropolis
MC method in Eqs.~(\ref{eq:nine}) and (\ref{eq:ten}). Performing a
term-wise comparison and omitting $O(R^3)$ and higher order terms, we found that
there is a one-to-one mapping between all FP terms of MC and LLG if:
%
%   Eq: Time quantification equation and precessional step size
%
\begin{eqnarray}
R^2\Delta\tau_{\mathrm{\scriptscriptstyle
MC}}&=&\frac{40\alpha}{1+\alpha^2}\frac{\gamma_0}{\beta\mu_{0}VM_s}\Delta
t_{\mathrm{\scriptscriptstyle LLG}} \label{eq:eleven}
\\
\Phi&=&\frac{\beta K_{u}V}{10\cdot\alpha}R^2 \label{eq:twelve}
\end{eqnarray}

\begin{figure}
\includegraphics[width=.43\textwidth, bb={13 13 260 200}]{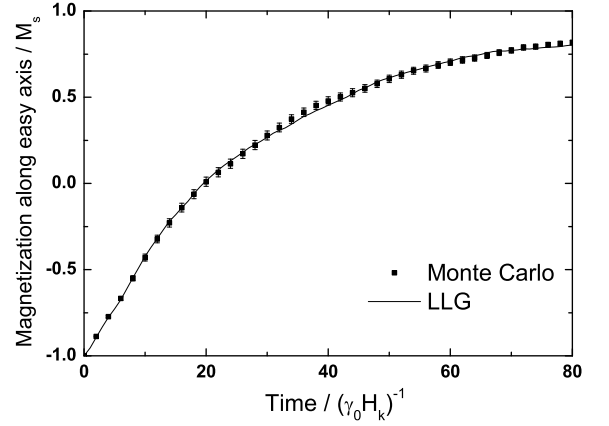}% Here is how to import EPS art
\caption{\label{fig:single} Time dependence of magnetization along
easy axis, for an isolated particle. $K_{u}V/k_{B}T=15$, applied
field $h=0.42$ tilted at $\pi/4$ relative to easy axis. Damping
constant $\alpha=0.5$.}
\end{figure}
\begin{figure}
\includegraphics[width=.43\textwidth, bb={13 13 260 200}]{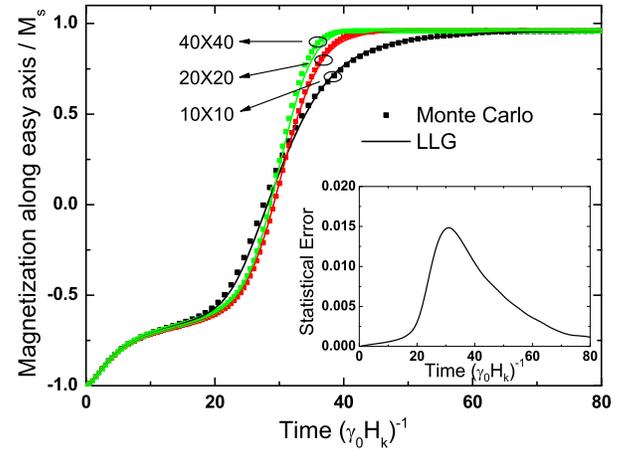}% Here is how to import EPS art
\caption{\label{fig:array} Time dependence of magnetization along
easy axis for interacting spin array. Periodic boundary conditions
were used and $K_{u}V/k_{B}T=25$, applied field $h=0.5$ at a tilted
angle of $\pi/4$ relative to the easy axis. Damping constant
$\alpha=1$, exchange coupling strength $J/K_u=2$ (Hamiltonian of an
interacting system with exchange coupling strength $J$ can be found,
i.e. in Ref.~\cite{hinzke}). $R = 0.025$ is used in the Monte Carlo
simulation. Statistical error for the $10\times10$ lattice Monte
Carlo simulation is shown in the inset.}
\end{figure}
Note that $\Phi$ is in order of $R^2$, thus we are justified in
neglecting $O(\Phi^2)$ terms in the above comparison between
Eq.~(\ref{eq:three}), and Eqs.~(\ref{eq:nine}) and (\ref{eq:ten}).
From Eq.~(\ref{eq:eleven}), we obtain the time quantification factor
of our hybrid Metropolis MC method, while Eq.~(\ref{eq:twelve})
determines the precessional step size $\Phi$. After taking into consideration the probability factor $q$, Eqs.~(\ref{eq:eleven}) and (\ref{eq:twelve}) can be reduced to Nowak's results \cite{nowak} in the high damping case.

To test the validity of Eqs.~(\ref{eq:eleven}) and
(\ref{eq:twelve}), we perform numerical calculations of the
switching process for a magnetic particle in which the easy axis is
oriented at $\pi/4$ to the applied field direction. All results are
averaged from a few thousand simulations. We consider the time
evolution behavior of the mean magnetization component along the
easy axis, and found a close convergence between our time-quantified
MC method and the LLG equation (Fig.~\ref{fig:single}). In these
calculations, we use $R=0.03$ for MC, and $\Delta t=0.001$ for the
LLG integration. We also apply our analytic results to $10\times10$,
$20\times20$ and $40\times40$ interacting spin array systems. For
these simulations, $R=0.025$ is used. We remark that a smaller step
size $R$ reduces statistical errors. We obtain a very good
convergence between the MC and LLG results for all three arrays
(Fig.~\ref{fig:array}), especially so for the larger arrays. We
believe this is due to the effects of self averaging.

\begin{figure}
\includegraphics[width=0.43\textwidth, bb={13 13 260 200}]{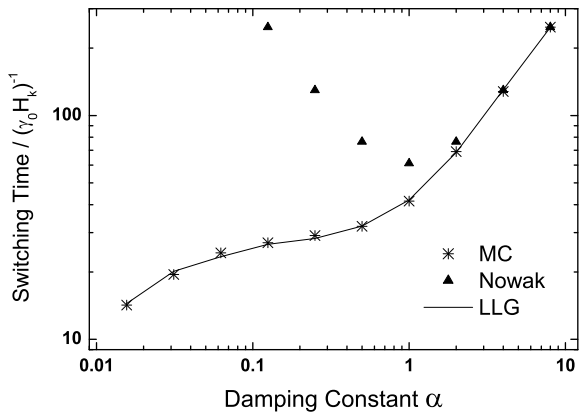}% Here is how to import EPS art
\caption{\label{fig:alpha} Switching time v.s. damping constant
$\alpha$. $K_uV/k_BT=15$, applied field $h=0.42$ at a tilted angle
of $\pi/4$ relative to easy axis. Errorbars are smaller than the
size of the symbols. Note that Nowak's method diverges from LLG
equation at $\alpha<2$.}
\end{figure}

Note that our derivation of the FPE factors is applicable in a very
general case. For instance, we do not require assumption of the
system being in the vicinity of an energy minimum \cite{nowak}. The
derivation also provides additional information, e.g., it explains
mathematically why the Metropolis MC random walk method of
Ref.~\cite{nowak} fails to include the energy conservative
precessional motion. The FPE expression for the pure Metropolis MC
does not contain terms corresponding to the $g'$-factor related
terms of the LLG method [Eq.~(\ref{eq:three})], which are precisely
the terms which reflect the precessional part of the magnetization
dynamics \cite{coffey}. Thus, as shown in Fig.~\ref{fig:alpha}, we
have successfully implemented the representation of precessional
motion in our MC method. We investigate the influence of damping
constant on switching time, where the switching time is defined as
the time required for the magnetic moment to reach zero from the
initial state. The precessional step size $\Phi$ guarantees a
precise description of switching process even in the case of very
low damping constant $\alpha$, in which precessional motion
dominates the reversal process \cite{ourjap}. By contrast, the
results obtained from the pure Metropolis MC method of Nowak
\emph{et al.} diverges significantly from that of the LLG equation
at low $\alpha$.

To summarize, we have proposed a general method using FPE to map MC
to LLG and vice versa. We derived the drift and diffusion FP terms,
corresponding to the MC method and compare them to the FP terms
obtained from the LLG dynamical equation. By matching the terms in
the drift and diffusion coefficients, we obtain a time
quantification factor and the required precessional step size. The
idea of using the FPE to link two stochastic methods (MC and
Langevin methods) is general, and may be applied to other areas such
as molecular dynamics in our future work.

This work was supported by the National University of Singapore Grant No.(R-263-000-329-112). Two of the authors H. K. Lee and Y. Okabe are partly supported by a Grant-in-Aid for Scientific Research from the Japan Society for the Promotion of Science.
%
%\bibliography{precessional}

\end{document}